\documentstyle[12pt]{article}

\begin{document}


\renewcommand{\thefootnote}{\fnsymbol{footnote}}

\begin{center}

\hfill KEK-TH-434\\
\hfill April 1995\\
\vskip .5in

{\large \bf String Field Theories from One Matrix Models}
\vskip .8in

{\bf Michio Ikehara\footnote{E-mail address:
ikehara@theory.kek.jp}}
\vskip .3in

{\em KEK theory group, Tsukuba, Ibaraki 305, Japan}\\
and\\
{\em Department of Physics, University of Tokyo,
Bunkyo-ku, Tokyo 113, Japan}\\
\vskip 1.1in

\end{center}

\begin{abstract}
Through the continuum limit of the one matrix model on
the multicritical point the corresponding Schwinger-Dyson equation
of temporal-gauge string field theory is derived.
It agrees with that of the background independent formulation
recently proposed.
\end{abstract}

\renewcommand{\thefootnote}{\arabic{footnote}}
\setcounter{footnote}{0}

\newpage


Discrete methods, realised analytically by matrix models,
have been instrumented in furthering
our understanding of string theory.
For example, it is known that
the double scaling limit\cite{DSL}
of the one matrix model on the $k$-th critical point\cite{kaz}
corresponds to a $(2,2k-1)$ type string\cite{lee}.
Recently important progress has been made in this direction
via the formulation of temporal-gauge string field theory
in the $c=0$\cite{ik0} and $c=1-6/m(m+1)$\cite{ik2,iikmns} cases.
The corresponding $W$ constraints\cite{fkn} expected
from matrix models can be derived from the appropriate
continuum Schwinger-Dyson equations.
This approach is motivated by the transfer matrix formalism
outlined in \cite{kkmw} which was later applied to the multicritical
case\cite{gk}.
These theories and multicritical points of the one matrix model
were studied via a background independent formulation in \cite{ik}.
These considerations were extended to closed and open strings
in \cite{mo} and \cite{ko}.

That such models are indeed connected with string theory
was verified when dynamical triangulation
was shown to yield $c=0$ string field theory
in the continuum limit\cite{W}.
Stochastic quantization of the matrix model
was also considered in \cite{jr}.
In addition to these studies the the continuum Schwinger-Dyson
equations for $c=0$ and $c=1/2$
were derived from the one and two matrix $\phi^3$ model,
respectively\cite{cont}.
In this letter we wish to use the techniques of \cite{cont}
to derive the continuum Schwinger-Dyson equation of the $k$-th
critical point of the one matrix model and confirm the results
first presented in \cite{ik}.

Let us consider the one matrix model.
We take the matrix $\phi$ to be an $N\times N$ hermitian matrix.
In order to obtain the $k$-th critical point
we define the action as
\begin{equation}
S(\phi)=N\mbox{tr}\left(\sum_{n=2}^{k+1}
{\lambda_n\over n}\phi^n\right).
\end{equation}
By rescaling $\phi$ we choose to set $\lambda_2=1$.
The loop operator of length $n$ is defined as
\begin{equation}
W(n)=
\left\{ \begin{array}{ll}
\displaystyle{{1\over N}\mbox{tr}\phi^{n}}& (n\geq 0)\\
\noalign{\vskip 1.2ex}
0 & (n \leq -1).
\end{array}\right.
\label{wn}
\end{equation}
With this operator we define the partition function of
loop amplitudes in the matrix model to be
\begin{equation}
Z_m(J)={\int d\phi e^{-S-S_J}\over \int d\phi e^{-S}},
\end{equation}
where
\begin{equation}
S_J(\phi)=-\sum_n J(n)W(n).
\end{equation}
In this approach the Schwinger-Dyson equation is expressed
by setting the integration of the total derivative
to be zero:
\begin{equation}
\int d\phi {1\over N^2}\mbox{tr}\left(
{\partial\over \partial \phi}
\phi^{n-1}\right)e^{-S-S_J}=0.
\end{equation}
Here $\partial/\partial\phi$ operates not only on $\phi^{n-1}$
but also on $S$ and $S_J$.
When $n\geq 1$ this equation can be expressed in terms of $Z_m(J)$.
Using the function $\theta(n)$ defined as
\begin{equation}
\theta(n)=\left\{
\begin{array}{ll}
1&(n\geq 1)\\
0&(n\leq 0),
\end{array}\right.
\end{equation}
we can rewrite the above equation as
\begin{eqnarray}
&&
\left[\sum_{m=-\infty}^\infty{\partial^2\over
\partial J(m)\partial J(n-2-m)}
+{\theta(n)\over N^2}\sum_{m=1}^\infty mJ(m)
{\partial\over\partial J(n-2+m)}\right.
\nonumber\\&&
\left.\mbox{}-\sum_{m=2}^{k+1}\lambda_m
{\partial\over\partial J(n-2+m)}
+\sum_{i=0}^{k-1}\delta_{n+i,0}\sum_{m=i+2}^{k+1}\lambda_m
{\partial\over\partial J(-i-2+m)}\right]Z_m(J)=0.
\label{sd1}
\end{eqnarray}
In the continuum limit the $\delta$-function and its
derivatives, expressed in terms of the loop length,
will appear in $W(n)$.
In order to obtain the continuum loop operator
we have to subtract these terms from $W(n)$.
We therefore employ the following partition function $Z_c$:
\begin{equation}
Z_m(J)=\exp\left(\sum_n J(n)c(n)\right)Z_c(J),
\end{equation}
where $c(n)$ consists of Kronecker $\delta$'s.
This $c(n)$ can be obtained by demanding
that the linear terms of $\partial/\partial J$
vanish in eq.(\ref{sd1}) expressed now via $Z_c$.
Therefore
\begin{equation}
c(n)=-{1\over 2}\sum_{m=2}^{k+1}\lambda_m\delta_{n+m,0},
\end{equation}
from which eq.(\ref{sd1}) becomes
\begin{eqnarray}
&&
\left[\sum_{m=-\infty}^\infty{\partial^2\over
\partial J(m)\partial J(n-2-m)}
+{\theta(n)\over N^2}\sum_{m=1}^\infty mJ(m)
{\partial\over\partial J(n-2+m)}\right.
\nonumber\\&&
\left.\mbox{}
-\sum_{m=-\infty}^{\infty}c(m)c(n-2-m)
+\sum_{i=0}^{k-1}\delta_{n+i,0}\sum_{m=i+2}^{k+1}\lambda_m
{\partial\over\partial J(-i-2+m)}\right]Z_c(J)
\nonumber\\&&
=0.
\label{sd2}
\end{eqnarray}
Note that since $W(0)=1$ we can replace $\partial/\partial J(0)$
in the last term of eq.(\ref{sd2}) by 1.
For simplicity we choose to drop $\partial/\partial J(m)$
when $m\geq 1$ in the last term by multiplying eq.(\ref{sd2})
by $n(n+1)\cdots (n+k-2)$.
We further define
\begin{equation}
J_c(n)=y_c^{-n}J(n),
\end{equation}
so that the loop operator becomes $y_c^n(W(n)-c(n))$.
Therefore eq.(\ref{sd2}) can be rewritten as
\begin{eqnarray}
&&
n(n+1)\cdots (n+k-2)\left[\sum_{m=-\infty}^\infty{\partial^2\over
\partial J_c(m)\partial J_c(n-2-m)}
+{\theta(n)\over N^2}\sum_{m=1}^\infty mJ_c(m)
{\partial\over\partial J_c(n-2+m)}\right.
\nonumber\\&&
\left.\mbox{}
-y_c^{n-2}\sum_{m=-\infty}^{\infty}c(m)c(n-2-m)
+y_c^{n-2}\lambda_{k+1}\delta_{n+k-1,0}\right]Z_c(J_c)=0.
\label{sd3}
\end{eqnarray}

In the continuum limit we let the length of
each side of the loop tend to zero, $a\rightarrow 0$,
with the loop length held fixed, $l=na$.
In this limit we have
\begin{equation}
\sum_n\rightarrow {1\over a}\int dl,
\end{equation}
\begin{equation}
\theta(n)\rightarrow \Theta(l),
\end{equation}
\begin{equation}
\delta_{n+i,0}\rightarrow a\delta(l+ia).
\end{equation}
Let us find the $k$-th critical point of this model.
Now that there are $k$ parameters, $y_c$, $\lambda_3$, $\lambda_4$,
$\cdots$, $\lambda_{k+1}$, we can choose the critical values of
these so that when we expand the last two terms
of eq.(\ref{sd3}) in $a$ each coefficitent
of $a$, $a^2$, $\cdots$, $a^k$ will vanish.
Since the last two terms of eq.(\ref{sd3}) are of order $a^{k+1}$
on this critical point
and eq.(\ref{sd3}) holds for every order of $a$,
the first two terms should also be of order $a^{k+1}$.
Therefore from the first term we find
\begin{equation}
{\partial\over\partial J_c(n)}\sim a^{k+{1\over 2}}
{\delta\over\delta j(l)},
\label{ddj}
\end{equation}
where $\delta/\delta j(l)$ creates the continuum loop operator:
\begin{equation}
Z_c(J_c)\rightarrow \left\langle\exp\left(\int_0^\infty
dlj(l)w(l)\right)\right\rangle\equiv Z[j].
\end{equation}
Eq.(\ref{ddj}) means that
\begin{equation}
J_c(n)\sim a^{-k+{1\over 2}}j(l),
\end{equation}
and in order to make the second term of eq.(\ref{sd3})
of the same order as the other terms we set
\begin{equation}
{1\over N^2}=a^{2k+1}g,
\end{equation}
where $g$ is the string coupling constant\cite{DSL}.
The order $a^{k+1}$ contribution of eq.(\ref{sd3}) then yields
the continuum Schwinger-Dyson equation:
\begin{equation}
\left[l^{k-1}\left\{\int_0^l dl'
{\delta^2\over\delta j(l')\delta j(l-l')}
+g\int_0^\infty dl'l'j(l'){\delta\over\delta j(l+l')}\right\}
+C\delta^{(k)}(l)\right]Z[j]=0,
\label{sd4}
\end{equation}
where $l\geq0$ and $C$ is some constant.
Note that there is a factor of $l^{k-1}$ which admits
the $k$-th critical point\cite{ik}.
This equation is the one on the critical point.
If the parameters approach the critical point as
\begin{equation}
\lambda_m=\lambda_m^{critical}\left(1+\sum_{n=2}^kA_{m,n}a^n\right),
\end{equation}
and if we choose $A_{m,n}$ so that the coefficient of
$a^3$, $a^4$, $\cdots$, $a^k$ in the last two terms
of eq.(\ref{sd3}) vanish,
then the terms proportional to $\delta^{(k-2)}(l)$,
$\delta^{(k-3)}(l)$, $\cdots$ will be added to eq.(\ref{sd4}).
By further choosing $A_{m,n}$ appropriately this equation will
become the case of non-zero cosmological constant.
If we drop the factor of $l^{k-1}$ in eq.(\ref{sd4})
more terms proportional to the $\delta$-function or
its derivative will appear.
The terms of this kind have already been derived in \cite{ik}.
They can be determined by requiring
that the continuum Schwinger-Dyson equation
becomes the Virasoro constraints with shifted variables.

In this letter we have shown that the $k$-th critical point of
the one matrix model becomes string field theory
whose Schwinger-Dyson equation have the factor of $l^{k-1}$.
This agrees with \cite{ik}.
If we drop this factor we will have to deal with
terms like $\delta_{n,0}\partial/\partial J(m)$
which were not considered here.
This kind of analysis may be necessary for higher critical points
of the two matrix model which correspond
to the $c=1-6/m(m+1)$ string.

\vskip 1cm

\begin{center}
{\bf Acknowledgement}
\end{center}

The author would like to thank H. Kawai and N. Ishibashi for
useful discussions and B. Hanlon for carefully reading the manuscript.



\begin{thebibliography}{99}

\newcommand{\NPB}{Nucl. Phys. B}
\newcommand{\PLB}{Phys. Lett. B}
\newcommand{\MPLA}{Mod. Phys. Lett. A}
\newcommand{\PRD}{Phys. Rev. D}
\newcommand{\PRL}{Phys. Rev. Lett. }
\newcommand{\CMP}{Commun. Math. Phys. }

\bibitem{DSL} E. Br\'{e}zin and V. Kazakov, \PLB236 (1990) 144;\\
              M. Douglas and S. Shenker, \NPB335 (1990) 635;\\
              D. Gross and A. Migdal, \PRL64 (1990) 127; \NPB340 (1990) 333.

\bibitem{kaz} V. Kazakov, \MPLA4 (1989) 2125.

\bibitem{lee} M. Staudacher, \NPB336 (1990) 349,\\
              E. Br\'{e}zin, M. Douglas, V. Kazakov and S. Shenker,
              \PLB237 (1990) 43,\\
              D. Gross and M. Migdal, \PRL64 (1990) 196,\\
              \u{C}. Crnkovi\'{c}, P. Ginsparg and G. Moore,
              \PLB237 (1990) 196.

\bibitem{ik0} N. Ishibashi and H. Kawai, \PLB314 (1993) 190.

\bibitem{ik2} N. Ishibashi and H. Kawai, \PLB322 (1994) 67.

\bibitem{iikmns} M. Ikehara, N. Ishibashi, H. Kawai, T. Mogami, R. Nakayama
                 and N. Sasakura, \PRD50 (1994) 7467.

\bibitem{fkn} M. Fukuma, H. Kawai and R. Nakayama, Int. J. Mod. Phys. A6
                 (1991) 1385;\\
              R. Dijkgraaf, E. Verlinde and H. Verlinde, \NPB348 (1991) 435.

\bibitem{kkmw} H. Kawai, N. Kawamoto, T. Mogami and Y. Watabiki,
               \PLB306 (1993) 19.

\bibitem{gk} S. S. Gubser and I. R. Klebanov, \NPB416 (1994) 827.

\bibitem{ik} N. Ishibashi and H. Kawai, KEK-TH-431, hep-th/9503134.

\bibitem{mo} T. Mogami, KEK-TH-426, hep-th/9412212.

\bibitem{ko} I. K. Kostov, SACLAY-SPHT-95-001, hep-th/9501135.

\bibitem{W} Y. Watabiki, INS-Rep.-1017, hep-th/9401096.

\bibitem{jr} A. Jevicki and J. P. Rodrigues, \NPB421 (1994) 278.

\bibitem{cont} M. Ikehara, \PLB348 (1995) 365.

\end{thebibliography}
\end{document}